# Is the Moon there if nobody looks: A reply to Gill and Lambare

Marian Kupczynski

Département de l'Informatique, UQO, Case postale 1250, succursale Hull, Gatineau. QC, Canada J8X 3X7

Correspondence: marian.kupczynski@uqo.ca

Keywords: Bell theorem, local realism, quantum entanglement, contextuality, Bell-CHSH inequalities, Bell Tests, local causality.

**Abstract:** In a recent preprint Gill and Lambare, criticize our paper published in Frontiers in Physics. Their criticism is unfounded and misleading. They define a probabilistic coupling, in which BI-CHSH hold for all finite samples. It does not mean, that BI-CHSH hold in our model, in which four incompatible experiments are described by setting dependent random variables implemented on 4 disjoint dedicated probability spaces. A joint probability distribution of these random variables does not exist and may not be used to derive inequalities. Moreover, their probabilistic coupling is useless, for a subsequent contextual model, which we construct to describe final data from Bell tests and to explain, in a locally causal way, the reported violations of inequalities and apparent violations of no-signaling. Neither quantum probabilistic model of an ideal EPRB experiment nor local realistic and stochastic hidden variable models may explain reported non-signaling Therefore; it is obvious that our model extends the set of probability distributions of possible measurements allowed in standard hidden variable models. Gill and Lambare, agree that Bell-CHSH inequalities are trivial arithmetic constraints for cyclic pairwise combinations of expectation values of jointly distributed random variables, which are never violated by finite sample. They also agree that pseudo random finite samples generated using local realistic and stochastic hidden variable models violate CHSH inequalities 50% of time, but not as often as in Bell Tests However, they seem not understand , the main message of our paper, that the violation of BI-CHSH and Eberhard inequalities by finite samples in Bell Tests, no matter how well these tests are designed and performed, does not allow for doubt regarding the existence of objective external physical reality and causal locality in Nature. Measurement outcomes in quantum domain are locally created in interaction of physical systems with measuring instruments in well-defined experimental contexts. This is why, we incorporated in a probabilistic model contextual setting dependent variables describing measuring instruments. Our contextual model does not want to circumvent Bell Theorem. Therefore the title of Gill and Lambare paper and their conclusion:" Kupczynski's escape route for local realism is not available" are misleading and have nothing to do with the content and conclusions of our paper.

## 1   Introduction

In their recent preprint [1], Gill and Lambare (G-L) criticize our paper *Is the Moon there if nobody looks: Bell inequalities and Physical Reality* [2] published in Frontiers in

Physics. They wrote: *the work is built around a mathematical claim by MK which is actually false, and MK's reasoning around this claimed assertion is also false.* They also seem to suggest, that their criticism applies to preceding papers [3-5] and even to the papers by other authors.

This statement and several other statements in [1] are misleading and incorrect. Everybody who takes time to read our paper(s) will agree. Our model is contextual and not local realistic. We do not want to circumvent Bell Theorem. We demonstrate that violation of inequalities and apparent violation of no-signalling in Bell Tests may be explained in a locally causal way without evoking quantum magic. Therefore, the violation of inequalities does not allow for doubt regarding the existence of objective external physical reality and causal locality in Nature

Joint probability distribution (JP) of N random variables describes random experiments in which N outcomes are outputted in each trial. In an ideal EPRB experiment and in Bell tests there is no JP of random variables $A_x$, $A_{x'}$, $B_y$ and $B_{y'}$ assigning ±1 to clicks on distant detectors. Expectations $E(A_x A_{x'} B_y B_{y'})$ do not exist and may not be used to derive inequalities. In our model, experimental raw data obtained in incompatible experimental settings (x, y) are described using dedicated probability spaces $\Lambda_{xy}$ and as Larsson and Gill demonstrated [6] :

$$|S| = |E(A_x B_y | \Lambda_{xy}) - E(A_x B_{y'} | \Lambda_{xy'})| + |E(A_{x'} B_y | \Lambda_{x'y}) + E(A_{x'} B_{y'} | \Lambda_{x'y'})| \leq 4 - 2\delta \quad (1)$$

In our model δ= 0, thus $|S| \leq 4$. CHSH inequalities $|S| \leq 2$, <u>may not be derived</u> and , random variables are taking values ±1 or 0. Our model describes only raw data in Bell Tests, which contain a lot of items (0,0), (a,0) or (b,0), thus estimated pairwise expectations and S are in general close to 0. This model is only used to derive a correct contextual description of the final data in Bell Tests.

G-L define a "toy probabilistic model" describing a random experiment using an experimental protocol, which may not be implemented in Bell Tests. In their model , in each trial values ($\lambda_1$, $\lambda_2$, $\lambda_x$, $\lambda_{x'}$, $\lambda_y$, $\lambda_{y'}$) of 6 hidden variables are generated and 4 values ($A_x(\lambda_1, \lambda_x)$, $A_{x'}(\lambda_1, \lambda_{x'})$, $B_y(\lambda_2, \lambda_y)$, $B_{y'}(\lambda_2, \lambda_{y'})$) are calculated and outputted. Random variables **A**x, **A**x′, **B**y and **B**y' are of course jointly distributed on a unique probability space $\Lambda$ and BI-CHSH can be rigorously derived. All pseudo-random finite samples generated using their model, may never violate these inequalities.

In Bell Tests and in our model: <u>a *joint probability distribution of all possible hidden events ($\lambda_x$, $\lambda_1$, $\lambda_y$, $\lambda_2$, $\lambda_{x'}$, $\lambda_{y'}$, $\lambda_2$) does not exist*</u> , because hidden events ($\lambda_x$, $\lambda_{x'}$) and ($\lambda_y$, $\lambda_{y'}$) may never occur together. This is why, one may not prove CHSH assuming the existence of such probability distribution and a non-vanishing $E(A_x A_{x'} B_y B_{y'})$.[2]

Using a precise current terminology, G-L define a probabilistic coupling for our model:

$$E(A_x) = E(\mathbf{A}_x) ,... E(B_{y'}) = E(\mathbf{B}_{y'}) \; ; \; E(A_x B_y) = E(\mathbf{A}_x \mathbf{B}_y),... E(A_{x'} B_{y'}) = E(\mathbf{A}_{x'} \mathbf{B}_{y'}) \quad (2)$$

It does not mean that CHSH inequalities hold in our framework. CHSH inequalities hold only in a probabilistic model describing a random experiment, in which 4 outcomes are outputted in each trial and may be displayed in Nx4 spreadsheet. Such protocol is neither consistent with the protocol of an ideal EPRB experiment nor with the protocol of Bell Tests. In 1972, De la Peña, Cetto and Brody [7] tried to explain it, in different words, but Bell clearly misunderstood their paper. They also strongly suggested, that hidden variables should depend on experimental settings. In 1982, Fine explained in detail, the relation between the existence of a JP and BI-CHSH inequalities [8].

Members of so called probabilistic opposition, who are indirectly criticised by G-L in [1] and by Lambare and Franco in [9], arrived ,often independently, to the same <u>correct</u> conclusions.

## 2     Finite samples and experimental protocols

Many different metaphysical arguments may justify a choice of a specific probabilistic model. However, once a probabilistic model is defined, its meaning and implications should be discussed rigorously using the language of mathematical statistics. A hidden variable $\lambda$ is a value of some hidden variable L and it cannot be a quantum mechanical wave function, as several authors claim.

As we discuss in detail in [10], Bell also proposed a non-contextual probabilistic coupling and proved that this coupling is inconsistent with the probabilistic model provided by quantum mechanics for the EPRB experiment.

As we explained in the introduction , the existence of a probabilistic coupling, does not prove, that CHSH inequalities hold in incompatible random experiments described by pairwise measurable random variables. Probabilistic coupling allows only, to find probabilistic bounds on how often and how seriously CHSH inequalities may be violated by finite samples in these experiments. For example, Gill correctly conjectured [11]:

$$\Pr\left(\langle AB \rangle_{obs} + \langle AB' \rangle_{obs} + \langle A'B \rangle_{obs} - \langle A'B' \rangle_{obs} \geq 2\right) \leq \frac{1}{2} \qquad (2)$$

where $\langle AB \rangle_{obs}$ is an estimate of E(AB) etc. More detailed critical discussion of various finite sample proofs of Bell-type inequalities may be found in [5].

There is a significant difference between a probabilistic model and a hidden variable model. If we average out some variables in a probabilistic model we obtain always a marginal probability distribution describing some <u>feasible</u> experiment. If we average out some variables in a hidden variable model we may obtain a new hidden variable model, which does not correspond to any feasible experiment.

There is a subtle relationship of probabilistic models with experimental protocols which we discussed in [12, 13]. As we demonstrated with Hans De Raedt, different experimental protocols, based on the same probabilistic model, may generate dramatically different estimates of various population parameters [14].

In the appendix of their paper, G-L reproduce, from [2], the proof of CHSH inequalities, given by Bell in 1971 [15]. After averaging over $\lambda_x$ and $\lambda_y$, one obtains a new hidden variable model describing a random experiment outputting 4 outcomes $\overline{A}_x(\lambda_1), \overline{A}_{x'}(\lambda_1), \overline{B}_y(\lambda_2), \overline{B}_{y'}(\lambda_2)$ in each trial. Then of course all finite samples obey strictly CHSH inequalities.

## 3 Locally causal description of Bell Tests

The discussion, whether the existence of a probabilistic coupling may be considered as a rigorous proof of CHSH inequalities for only pairwise measurable random variables <u>has no much importance</u>. CHSH are violated by pseudo-random samples generated using various probabilistic models. They are violated by experimental data in physics and in cognitive science. The important questions are:

1) Can we explain a statistical scatter of experimental outcomes in Bell Tests using a hidden variable probabilistic model consistent with local causality or not?

2) What metaphysical conclusions, if any, may be made, if the data violate significantly CHSH inequalities in a given experiment?

This is what we wanted to answer in [2]. The contextual model, so vigorously attacked by G-L describes only raw data in Bell Tests. Final data have to be extracted/post-selected from raw data in order to estimate pairwise expectations and to test CHSH inequalities. The final data are described by appropriate conditional pairwise expectations deduced using our model [4, 16, 17]. It does not mean that *Kupczynski is falling back on detection loophole*. Talking about detection loophole is misleading, because it suggests defective functioning of some measuring instruments, which may be avoided in future experiments. We explain below, why it may not be done. .

Raw data are obtained by conversion of two distant time- series of clicks into samples containing paired outcomes (a, b), with a=±1 or 0 and b== ±1 or 0, coding clicks in some synchronized time windows. From raw data, final data are extracted with only non-vanishing pairs (a, b) and pairwise expectations of random variables may be described as conditional expectations:

$$E(A_x B_y \mid A_x \neq 0, B_y \neq 0) = \sum_{\lambda \in \Lambda'_{xy}} A_x(\lambda_1, \lambda_x) B_y(\lambda_2, \lambda_y) p_x(\lambda_x) p_y(\lambda_y) p(\lambda_1, \lambda_2) \quad (3)$$

where $\Lambda'_{xy} = \{\lambda \in \Lambda_{xy} \mid A_x(\lambda_1, \lambda_x) \neq 0 \text{ and } B_y(\lambda_2, \lambda_y) \neq 0\}$. It explains, in a locally causal way, the apparent violation of no- signalling reported in [18-22]:

$$E(A_x | A_x B_y \neq 0) \neq E(A_x | A_x B_{y'} \neq 0); E(B_y | A_x B_y \neq 0) \neq E(B_y | A_{x'} B_y \neq 0) \quad (4)$$

Since final data in Bell Tests violate significantly non-signalling (3), we have to use Contextuality by Default (CbD) approach of Dhafarov and Kujala [23-25]. Random experiments performed in 4 incompatible pairs of settings are described now by 8 binary random variables: ($A_{xy}$ $B_{xy}$, … $A_{x'y'}$, $B_{x'y'}$) instead of 4. The final data in Bell Tests are described by a new explicitly contextual probabilistic mode [16]:

$$E(A_{xy} B_{xy}) = \sum_{\lambda \in \Lambda_{xy}} A_x(\lambda_1, \lambda_x) B_y(\lambda_2, \lambda_y) p_{xy}(\lambda_x, \lambda_y) p(\lambda_1, \lambda_2) \quad (5)$$

where $A_{xy} = \pm 1$ and $B_{xy} = \pm 1$. It is clear that, neither G-L coupling nor Bell averaging, over instrument variables may be used to prove CHSH inequalities for random experiments described by the probabilistic models (4) or (5).

The model (5) allows for more flexibility and more precise predictions, because variables describing distant measuring instruments may be correlated. The correlation does not mean causation, thus it does not imply spooky influences between them. Transmission probabilities between two polarisation filters obey the Malus law, which depends only on cos θ, where θ is a relative angle between polarisation axes of polarisation filters. Therefore, if λ are hidden variables describing a polarisation filter how it is "perceived" by an incoming beam at the moment of a measurement, it is plausible to assume, that after a rotation by an angle θ, the same filter is described by hidden variables λ'=f(λ, cos θ). Therefore one may try to explain θ dependence of estimated expectations E ($A_{xy} B_{xy}$), assuming that $p_{xy}(\lambda_x, \lambda_y)$ is also a function of cos θ The importance of rotational invariance was strongly advocated by Hess[26].

We agree that correlations between distant experimental outcomes may be called nonlocal. Nevertheless, they may be explained, in a locally causal way, using the models (4) or (5). The experimental protocol used in (4) is consistent with the experimental protocol used by Weihs et al. [27].

Delft experiment [28] used completely different experimental protocol based on so called "entanglement swapping". However, Delft experiment could not avoid using time-windows and post-selection [21, 22]. As we explained in [3] entanglement swapping may be also understood without evoking quantum magic. Contrary to what Aspect claimed: "*Mixing two photons on a beam splitter and detecting them in coincidence entangles the electron spins on the remote NV centers*" [29], the observation of a particular coincidence signal gives only the information, that "correlated signals" in distant laboratories were created and measurements were done in specific synchronized time slots.

# 4    Conclusions

There is no false mathematical claim and false assertions in our paper [2]. Our model is contextual and not local realistic. Signals arriving to measuring stations are described by setting independent random variables which are statistically dependent and causally independent. Measuring instruments are described by random variables which are setting dependent, they are causally independent but they may be statistically dependent (5). model (5) they are statistically dependent. We are not looking for escape route for local realism.

The existence of a non-contextual probabilistic coupling for 4 probabilistic models describing 4 incompatible random experiments does not mean, that CHSH inequalities hold for these experiments.

Talking about the efficiency, detection or coincidence loophole is misleading. If two time series of clicks are produced by <u>invisible</u> signals in distant laboratories, then to estimate correlations, we have to process data and construct generalised joint empirical frequency distributions [13]. One may rather talk, about *photon identification loophole* [30] , which can never be closed.

The hidden variables describing measuring instruments are explicitly incorporated in the models (4) and (5), thus they do not suffer from theoretical *contextually loophole* [30, 31].

The violation of inequalities and apparent violation of non-signalling in Bell Tests may be explained in a locally causal way without evoking quantum magic. Therefore, the violation of inequalities does not allow for doubt regarding the existence of objective external physical reality and causal locality in Nature.

Metaphysical conclusions of the violation of inequalities in Bell Tests are quite limited [2, 33]. The contextual character of quantum observables and active role played by measuring instruments was explained by Bohr many years ago. The violation of inequalities does not prove the completeness of QM, which was the subject of Bohr – Einstein quantum debate [3].

The speculations about *quantum nonlocality* and *quantum magic* are rooted in incorrect interpretations of QM and/or in incorrect "mental pictures" and models trying to explain invisible details of quantum phenomena [2, 34-36]. There is nothing magical in "nonlocal" correlations. They exist as well in classical physics and they may be explained by various conservation laws.

Nevertheless, the research stimulated by Bell papers and beautiful experiments designed and performed to test various inequalities paved the road to important applications of "nonlocal "quantum correlations in quantum information and in quantum technologies.